\documentstyle[12pt,epsfig]{article}
\input epsfig.sty
\input wsci.sty 
\begin{document}
\begin{flushright}
{\bf LPC 95-41/CONF}\\
{\bf hep-ph/9509390}\\
August 1995
\end{flushright}
\vspace*{0.6cm}

\centerline{\normalsize\bf CONSTRAINTS ON COMPOSITE-MODELS} 
\baselineskip=18pt 
\centerline{{\bf EFFECTIVE LAGRANGIANS FROM} 
\mbox{\boldmath $ 0\nu\beta\beta $} {\bf DECAY}.
\footnote{
Invited Talk
Presented at the International 
Workshop on Double Beta Decay, ECT$^*$, April 24 - May 5, 1995, Trento, 
Italy.}
\footnote{
Work partially supported by the EU program ``Human Capital and Mobility''
under contract No. CHRX-CT92-0026.}
}
\vspace*{0.6cm}
\centerline{\footnotesize O.~PANELLA\footnote{Presently at: Laboratoire
de Physique Corpusculaire, Coll\`ege de France, Paris, France.
E-mail: {\sf
panella@cdf.in2p3.fr}}}
\baselineskip=13pt
\centerline{\footnotesize\it Dipartimento di Fisica, Universit\`a di Perugia
and INFN, Sezione di Perugia}
\baselineskip=12pt
\centerline{\footnotesize\it Via A.~Pascoli, I-06100, Perugia, Italy}
\baselineskip=12pt

\centerline{\footnotesize E-mail: {\sf panella@perugia.infn.it}}

\vspace*{0.6cm}
\abstracts{
We give a brief review of the existing bounds on effective couplings
for excited states ($e^*, \nu^*, u^*, d^* $) 
of the ordinary quarks and leptons arising from a composite model scenario. 
We then explore the phenomenological
implications of the hypothesis that the excited neutrino $\nu^*$ might
be of Majorana type. Recent bounds on the half-life of the $\Delta L
=2 $ neutrinoless double beta decay ($0\nu\beta\beta $) 
are used to constraint the compositeness effective couplings. 
We show that the bounds so obtained
are roughly of the same order of magnitude  as those available 
from high energy experiments.
}
\normalsize\baselineskip=15pt
\setcounter{footnote}{0}
\renewcommand{\thefootnote}{\alph{footnote}}
\section{Introduction} 

\begin{figure}[htb]
\centering
\vspace*{4pt}
\mbox{\epsfig{file=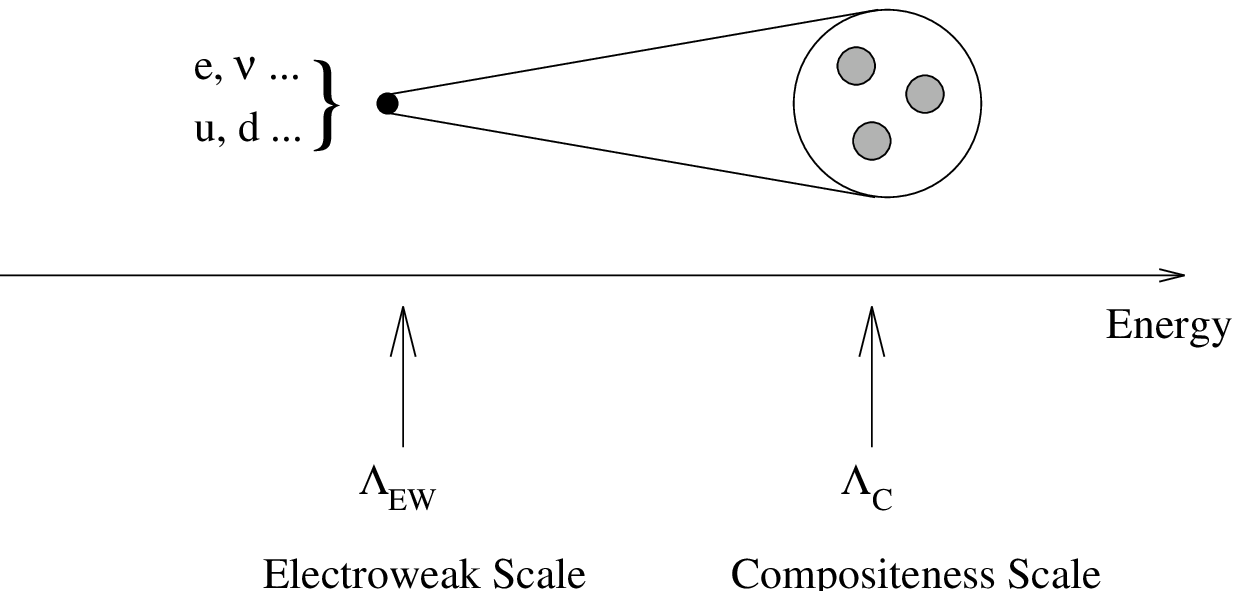,width=11.0cm}}
\vspace*{8pt}
\fcaption{
The idea of compositeness: at the energy scale 
$\Lambda_{\hbox{c}}$ (the compositeness scale, still unknown), 
ordinary quarks and leptons might show an internal structure. 
}
\label{scheme}
\end{figure}

Up to now the behaviour of quarks and leptons
has been, so far, very successfully described by
a theory of pointlike quantum fields interacting through 
$SU(2)\times U(1)\times SU(3)_c $ gauge interactions, which is 
nowadays referred to as the standard model. 
We may  however speculate that, when exploring higher energy ranges
(from $1$ TeV up to  $15 $ TeV 
as planned with the next generation of supercolliders
like LHC or NLC), we might hit an energy scale $\Lambda_{\hbox{c}}$ at 
which a sub-structure of those ``elementary'' particles will show up,
(see Fig.~\ref{scheme} for a schematic illustration).
Although so far there is no experimental evidence signaling
a further level of sub-structure, we cannot a priori exclude
this possibility, and must explore its phenomenological consequences.

The idea that quarks and leptons might not be genuine elementary
particles has been around for quite some time.
Many models
based on the idea that quarks and leptons are bound states
of still unknown entities (generally referred to as {\it preons}),
have already been proposed\cite{hara} but no completely consistent
dynamical theory has been found so far. 
It is therefore particularly important to study 
model independent features of the idea of compositeness. 
There are two main consequences of having composite
quarks and leptons: 
\begin{itemize}
\item[{(i)}]  four-fermion contact interactions of dimension six; 
\item[{(ii)}] highly massive excited states which couple 
to the ordinary fermions through
gauge interactions; 
\end{itemize}
Both facts are expected to give observable
deviations from the predictions of the standard model provided
that  the 
compositeness scale $\Lambda_{\hbox{c}}$ is not too large.
      
In this work we will first review what bounds on the compositeness
scale  can be derived from the study of the above mentioned
effective interactions, and then we will show how current lower 
bounds for the half-life 
of the neutrinoless double beta decay $(0\nu\beta\beta)$ can
be used to get constraints on the compositeness effective couplings,
when assuming the existence of a heavy composite Majorana 
neutrino.

\section{Contact Interactions} 

In preon models, modifications to the gauge boson propagators
and the interaction vertices with fermions are expected.
To describe the former one can use\cite{chano} a simple  
parametrization by multiplying  gauge  boson 
propagators by a form factor 
$F(Q^2) \approx 1+Q^2/{\Lambda_{\hbox{c}}^2} $.
Such form factors 
effects were experimentally searched for, already in 1981-82 at 
PETRA,
looking for deviations from the standard model 
predictions in the cross sections for $e^+ e^- \to f \bar{f} $
with $(f = e,\mu,\tau,q)$. These experiments~\cite{tasso} gave lower bounds 
on $\Lambda_{\hbox{c}}$ of the order of $100 $ GeV.

\begin{figure}[htb]
\vspace*{4pt}
\begin{center}
\mbox{\epsfig{file=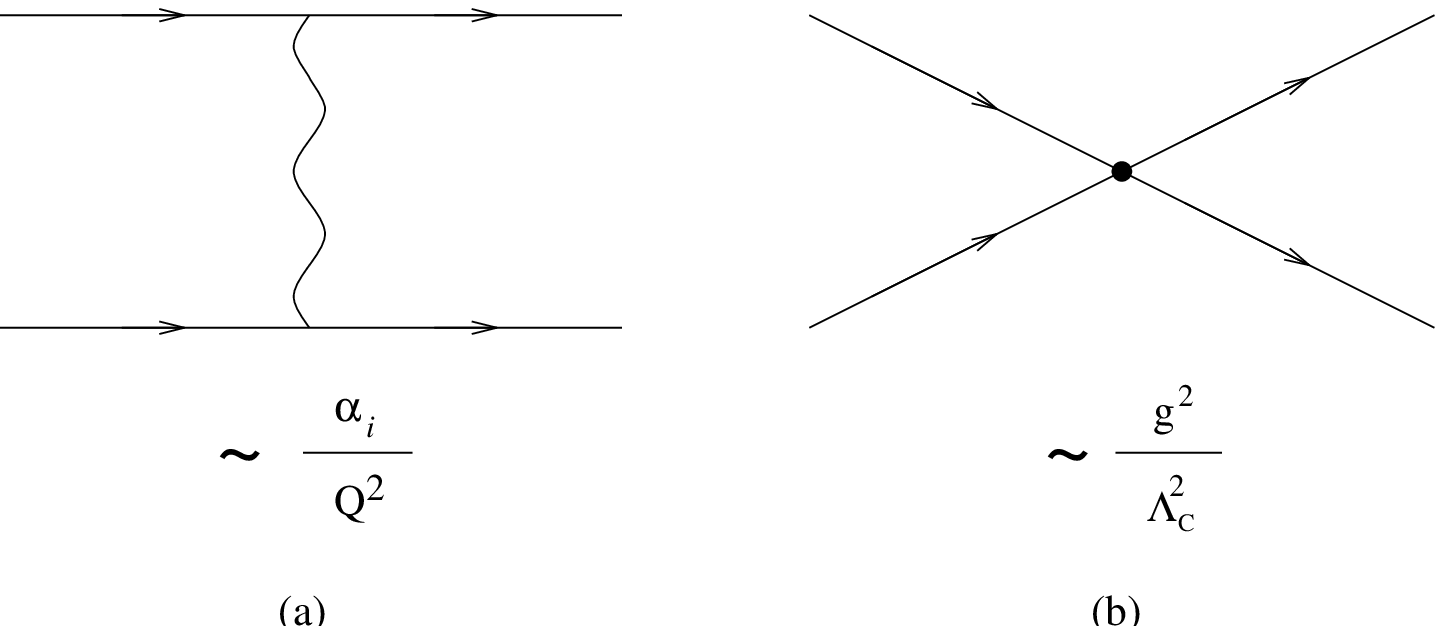,width=10.0cm}}
\end{center}
\vspace*{8pt}
\fcaption{
Contact interactions
give modifications  to the fermion fermion scattering amplitude. 
The interference between the standard model diagram (a) and the one coming
from contact interactions (b) will be the most important correction
relative to the standard model prediction.
}
\label{contact}
\end{figure}

Composite fermions are also expected to have additional
effective four fermions interactions through constituent
exchange. Eichten, Lane and Peskin\cite{eichten} proposed the
following effective lagrangian to parametrize
flavour diagonal helicity conserving contact interactions:   

\begin{equation}
{\cal L}_{ Cont.}^{(ff )} 
= \frac{g^2}{2\Lambda_{\hbox{c}}^2} 
\biggl[ \sum_{i,j=R,L} \eta_{ij}\, \bar{f}_j\gamma_\mu f_i \,
\bar{f}_j \gamma^\mu f_j 
\biggr]
\end{equation}
where the compositeness scale $\Lambda_{\hbox{c}}$ is defined 
in such a way that $g^2/4\pi = 1 $ (i.e. the coupling $g$ is strong) and 
$\hbox{max}{|\eta_{ij}| = 1 }$ and $i,j=L,R $.
In Fig.~\ref{contact} we show how contact interactions can 
affect fermion scattering at high energies. 
Clearly the interference between the contact term and the standard
model diagram will give contributions that, relative to the 
standard model one, will be of the order:
\begin{equation}
\simeq (\frac{\alpha_i}{Q^2})^{-2} 
\frac{\alpha_i g^2}{Q^2\Lambda_{\hbox{c}}^2}
= \frac{g^2}{\alpha_i}\frac{Q^2}{\Lambda_{\hbox{c}}}
\end{equation} 
and will thus overwhelm the form factor contributions which are
of the order $Q^2/\Lambda_{\hbox{c}}^2$.

The effective lagrangian in Eq. (1) can be extended to include
(model dependent) possibile non-diagonal contact
interactions (which are expected if different fermions share
common constituents). We could have:
\begin{equation}
{\cal L}_{cont.}^{(e\mu)} = \frac{g^2}{\Lambda_{\hbox{c}}^2}
\sum_{i,j=R,L} \eta_{ij} \bar{e}_i \gamma_\lambda e_i \bar{\mu}_j
\gamma^\lambda \mu_j
\end{equation}
and similarly for $(e,q)$.

Let us now discuss what bounds on the compositeness scale can be 
derived from the non-observation of deviations from the standard model
predictions.
We will use the following notation for the compositeness
scale:
$\Lambda_{ij}^{\pm}$ corresponds to the choice $\eta_{ij} = \pm 1$
and $\eta_{kl} = 0 $ for $k\neq i, l\neq j$.
High-energy fermion scattering, at electron positron and/or hadron 
colliders, has been used to study possible mani\-festations of the
effective lagrangians given in Eqs. (1,3). Recent bounds come
from the LEP experiment\cite{buskulic} 
at CERN where the study of the process
$e^+e^- \to e^+e^- (\mu^+ \mu^-)$ at the energy of the
$Z^0$ resonance has given lower bounds on the compositeness
scale: $ \Lambda_{LL}^+(eeee) > 1.6 $ TeV and 
$\Lambda_{LL}^+(ee\mu\mu) > 2.6 $ TeV. 

\begin{table}[t]
\begin{minipage}[b]{\textwidth}
\tcaption{
Lower bounds on the compositeness scale $\Lambda_{\hbox{c}}$
from high energy fermion scattering and from leptonic tau decays.
}\label{tab1}
\vspace{5pt}
\small
\centering
\begin{tabular}{||c|c|c||}\hline\hline
{} &{} &{}\\
process & $\Lambda_{LL}^+ (\hbox{TeV}) > $ 
& $\Lambda_{LL}^- (\hbox{TeV}) >$\\
{} &{} &{}\\
\hline
{} &{} &{}\\
$e^+ e^- \to e^+ e^- $ & 1.6 & 3.6 \\
{} &{} &{}\\
\hline
{} &{} &{}\\
$e^+ e^- \to \mu^+ \mu^- $ & 2.6 & 1.9 \\
{} &{} &{}\\
\hline
{} &{} &{}\\
$p \bar{p} \to e^+ e^- +X $ & 1.7 & 2.2 \\
{} &{} &{}\\
\hline
{} &{} &{}\\
$\tau \to \nu_{\tau} e \bar{\nu}_{e} $ & 3.8 & 8.1\\
{} &{} &{}\\
\hline\hline
\end{tabular} 
\end{minipage}
\end{table}

The Drell-Yan process ($p \bar{p} \to e^+ e^- + X $) has been 
used at FERMILAB to obtain lower bounds on the 
compositeness scale of contact interactions between quarks and 
leptons\cite{abe1}: 
$\Lambda_{LL}^+(eeqq) > 1.7 $ TeV.
Recently Diaz-Cruz and Sampayo\cite{diaz} derived bounds on 
$\Lambda(\tau\nu_{\tau} e\nu_e )$ from a theoretical analysis
of the effect of contact interactions in the leptonic $\tau$ decays,
as shown in Fig. 3.
\begin{figure}[htb]
\vspace*{4pt}
\begin{center}
\mbox{\epsfig{file=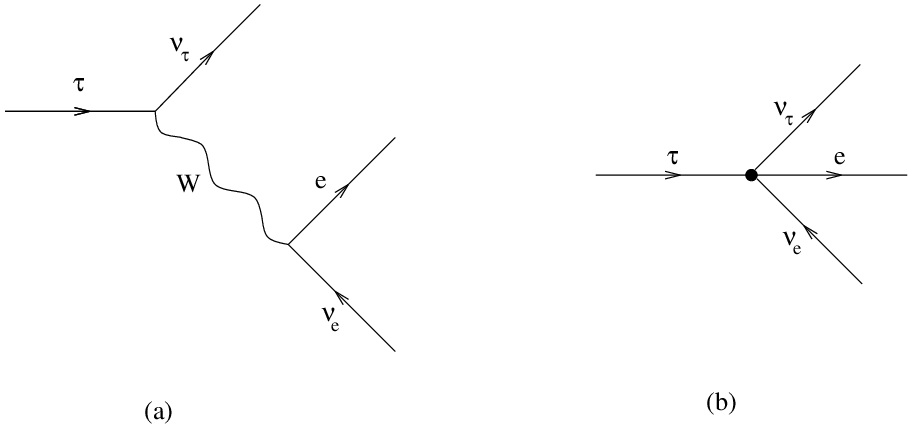,width=10.0cm}}
\end{center}
\vspace*{8pt}
\fcaption{
Leptonic $\tau$ decays: (a) standard model contribution; (b) contact
interactions contributuion.
}
\label{taudecay}
\end{figure}

They find a rather interesting bound; 
$\Lambda_{LL}^+(\tau\nu_\tau e\nu_e) > 3. 8 $ TeV 
but it is to be remarked 
that their result is based on the assumption that
the compositeness scale is flavour dependent, which leads to 
$\Lambda(\tau\nu_\tau e\nu_e) \ll \Lambda (\mu \nu_\mu e \nu_e) $.
The above mentioned 
bounds are summarized in Table 1. We also refer the reader
to the Review of Particle Properties of the Particle Data 
Group~\cite{pdg} for a list of previous bounds.

\section{Excited states of ordinary fermions}
Although no completely consistent dynamical composite theory 
has been found up to now,    
one inevitable common prediction of composite models 
is the existence of excited states of the known 
quarks and leptons, analogous to the  series of higher energy levels
of  the hydrogen atom.
The masses of the excited particles should not be much lower 
than the
compositeness scale $\Lambda_{\hbox{c}}$, which is expected to be at
least of the order of a TeV  according to the experimental 
constraints discussed in the previous section.
We expect therefore the heavy fermion masses to be, {\it at least},
of the order of a few hundred GeV. 

Excited leptons ($l^*$) and quarks $(q^*)$ are expected 
to interact with light fermions via gauge interactions.
Let us consider the coupling $e^* e \gamma $, depicted 
in Fig.~4 and for which  Low\cite{low}
proposed  in 1965 a magnetic moment type interaction:
\begin{equation}
{\cal L}_{int} = \frac{ef_\gamma}{2m^*} \bar{\psi}_{e^*}\sigma_{\mu\nu}
\psi_e F^{\mu\nu} + \hbox{h.c.}
\end{equation}
where $m^*$ is the mass of the excited electron and $f_\gamma$ is 
a dimensionless coupling constant; $F_{\mu\nu}$
is the electromagnetic field strength tensor.
\begin{figure}[htb]
\vspace*{4pt}
\begin{center}
\mbox{
\epsfig{file=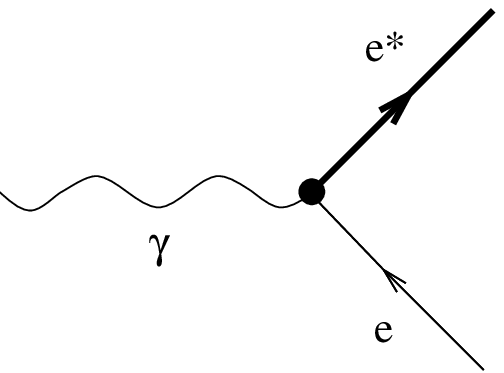,width=4.cm
}}
\end{center}
\vspace*{8pt}
\fcaption{
The transition coupling between a light electron to its
correspondent excited state ($e^*$) via gauge interactions.
}
\end{figure}

Clearly, this coupling can produce deviations from the predictions
of the standard model through the exchange of virtual heavy
excited states, or it may be responsible for the direct production and
decay of the excited states in high energy fermion scattering. 
These effects can be used to derive bounds on the coupling constants
appearing in Eq.~4.
In 1982 Renard~\cite{renard} showed that the precise
measurements of the anomalous electron magnetic moment give
bounds on the masses of the excited states (or equivalently
the compositeness scale). In Fig.~5 we show one of the 
diagrams involving the exchange of a virtual excited 
electron and contributing to the electron's anomalous magnetic moment and
electric dipole moment.
Renard considered a general tensor and pseudo-tensor effective
coupling:
\begin{equation}
{\cal L}_{int} =\frac{e}{2m^*}\bar{\psi}_{e^*}\, \sigma_{\mu\nu}
(a-ib\gamma_5)\psi_e \, F^{\mu\nu}+\hbox{h.c.}
\end{equation}
which is  a simple  generalization of Eq.~4.
\begin{figure}[htb]
\vspace*{4pt}
\begin{center}
\mbox{
\epsfig{file=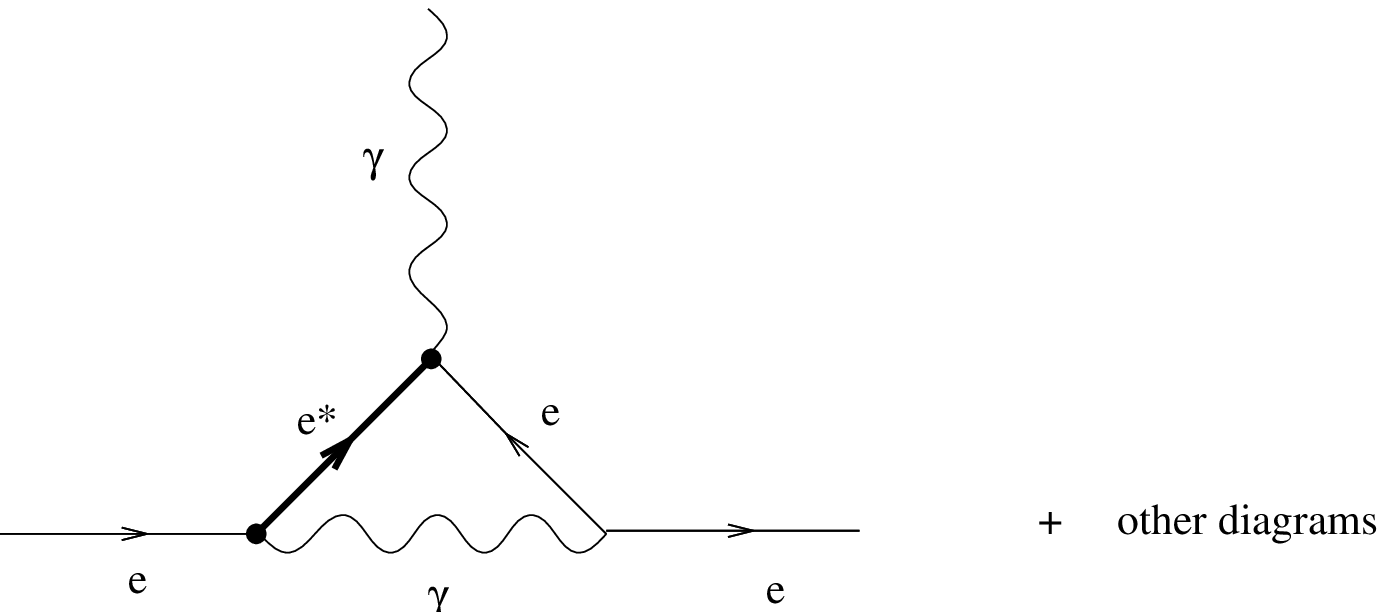
,width=10.cm
}}
\end{center}
\vspace*{8pt}
\fcaption{
One of the diagrams contributing to the electron anomalous magnetic
moment, in addition to those of the standard model, in the
hypothesis that excided states exist.
}
\end{figure}

Defining the anomalous magnetic moment and electric dipole moment
of the electron through the lagrangian:
\begin{equation}
{\cal L} = e \bar{\psi}_e 
\bigg\{ 
\gamma^\mu A_\mu +\frac{\chi}{4m_e}\sigma^{\mu\nu} F_{\mu\nu}
-i\frac{\chi'}{4m_e}\sigma^{\mu\nu}F_{\mu\nu}\gamma^5
\biggl\}
\bar{\psi}_e
\end{equation}
in the limit $m_{e^*} \gg m_e $ one obtains \cite{renard}:
\begin{eqnarray}
\chi & = & \frac{4\alpha}{\pi}(|a|^2 -|b|^2)\frac{m_e}{m_{e^*}} +
\frac{9\alpha}{2\pi}(|a|^2+|b|^2)\frac{m_e^2}{m_{e^*}^2}\, ; \cr
\chi' & = &\frac{8\alpha}{\pi}\Re (ab^*)\frac{m_e}{m_{e^*}}\, .
\end{eqnarray}

The small value of the electron dipole moment $d_e = \chi'/2m_e \simeq
0.7 \times 10^{-24}$ cm implies that $a$ and $b$ in Eq.~5 cannot be 
simultaneously real: $\Re (ab^*) \simeq 0$.
Regarding the anomalous magnetic moment contribution, we see that 
the effect appears at order $m_e/m_{e^*}$ if $|a|\neq |b|$, 
while if there is {\it chiral
symmetry} $(|a| =|b|)$ the effect is of order $(m_e/m_{e^*})^2$.
Thus, the precise measurement of the electron's $g-2$ will put 
a  weaker or stronger contraint on the value of $m_{e^*}$ 
depending on whether or not 
the coupling respects chiral symmetry.
From $\delta \chi_e \leq 2\times 10^{-10}$ one finds:
\begin{eqnarray}
m_{e^*} & \geq & (|a|^2 -|b|^2) \times 22\,  \hbox{TeV} \qquad 
\hbox{without chiral symmetry}\cr
m_{e^*} & \geq & (|a|^2 +|b|^2)^{1/2} \times 3.8\,  \hbox{GeV} \qquad 
\hbox{with chiral symmetry}
\end{eqnarray}
the corresponding bounds for the muon are:
\begin{eqnarray}
m_{\mu^*} & \geq & (|a|^2 -|b|^2) \times 110\,  \hbox{TeV} \qquad 
\hbox{without chiral symmetry}\cr
m_{\mu^*} & \geq & (|a|^2 +|b|^2)^{1/2} \times 110\,  \hbox{GeV} \qquad 
\hbox{with chiral symmetry}
\end{eqnarray}
One is thus led to conclude that if  the compositeness scale 
$\Lambda_{\hbox{c}}$ ($\approx m^*)$ is of the order of one to a few
TeV, then the coupling in Eq.~5 must display chiral symmetry, i.e.
the heavy excited state can couple only to a left-handed (or only to
a right-handed) light fermion.

The extension of the effective coupling  given in Eq.~5 including
electroweak interactions has been discussed in the  
literature~\cite{cab}   
using weak isospin ($I_W$) and hypercharge ($Y$) conservation.
Within this model, it is assumed that 
the lightness of the ordinary leptons could be related to some
global unbroken chiral symmetry which would produce massless 
bound states of preons in the absence of weak perturbations due to
$SU(2)\times U(1) $ gauge and Higgs interactions. 
The large mass of the excited leptons arises from the unknown underlying
dynamics and {\it not} from the Higgs mechanism. 

Assuming that such states are grouped in $SU(2) \times U(1)$ 
multiplets, since  light  fermions have $I_W=0,1/2$
and electroweak  gauge bosons have $I_W=0,1$,
only
multiplets with $I_W \leq 3/2$ can be excited in 
the lowest order in perturbation theory.   
Also, since none of the gauge fields carry hypercharge, a
given excited multiplet can couple only to a light multiplet with 
the same $Y$. 

In addition, conservation of the electromagnetic
current forces the transition coupling of heavy-to-light fermions
to be of the magnetic moment type respect to any electroweak
gauge bosons~\cite{cab}.
In fact, a $\gamma_{\mu}$ transition coupling between $e$ and  $e^*$
mediated by the ${\vec W}^{\mu}$ and $B^{\mu}$ gauge fields,
would result in an electromagnetic current of the type $j^{\mu}_{e.m.}
\approx \bar{\psi}_{e^*} \gamma^{\mu} \psi_e$ wich would not be 
conserved due to the different masses of excited and ordinary
fermions, (actually it is expected that $m_{e^*} \gg m_e$).

Let us here restrict to the first family and consider 
spin-$1/2$ excited states grouped in multiplets with 
$I_W=1/2 $ and $ Y=-1$, 
\begin{equation}
L^*= {\nu^* \choose e^*} 
\end{equation} 
which can couple to the light  left-handed multiplet
\begin{equation}
\ell_L = {\nu_L \choose e_L} ={{1-\gamma_5} \over 2}
{\nu \choose e}
\end{equation}
through the gauge fields ${\vec W}^{\mu}\, \hbox{and} B^{\mu}$.
The relevant interaction can be written\cite{cab} 
in terms of two {\it new} independent coupling constants $f$ and $f'$:
\begin{eqnarray}
{\cal L}_{int}& = & \frac{gf}{\Lambda_{\hbox{c}}} \bar{L}^*\sigma_{\mu\nu}
\frac{\vec\tau}{2} l_L \cdot \partial^\nu\vec{W}^\mu \cr
& & \phantom{xxxxx}
+\frac{g'f'}{\Lambda_{\hbox{c}}}
\biggl(-\frac{1}{2} \bar{L}^*\sigma_{\mu\nu}
l_L \biggr)\cdot \partial^\nu B^\mu + \hbox{h.c.}
\end{eqnarray}
where ${\vec \tau}$ are the Pauli $SU(2)$
matrices, $g$ and $g'$ are the usual $SU(2)$ and $U(1)$ gauge coupling
constants, and the factor of $-1/2$ in the second term is the
hypercharge of the $U(1)$ current. 
This effective lagrangian has been widely used in the literature
to predict production cross sections and decay rates of the excited
particles~\cite{pdg,ruj,baur}.

In terms of the physical gauge fields 
$Z_\mu= - B_\mu\sin\theta_W \ + W^3_\mu\cos\theta_W$,\\ 
$A_\mu = B_\mu \cos\theta_W +W^3_\mu\sin\theta_W $ and
$W^{\pm}_{\mu}=(1 / \sqrt 2) \Bigl( W^1_{\mu} \mp \, i \, 
W^2_{\mu} \Bigr)$, 
the effective interaction in Eq.~12 can be reexpressed as
\begin{equation}
{\cal L}_{int} = \sum_{V=\gamma,Z,W} \frac{e}{\Lambda_{\hbox{c}}}
\bar{f}^*\, \sigma_{\mu\nu}(c_{Vf^*f}-d_{Vf^*f}\gamma_5) \, f\,  
\partial_\mu V_\nu +\hbox{h.c.}
\end{equation}
where the coupling constants have to satisfy the condition 
$|c_{Vf^*f}| = |d_{Vf^*f}|$, if we require chiral symmetry, and are 
related to $f$ and $f'$ by the following relations:
\begin{eqnarray}
c_{\gamma f^*f} & = & -\frac{1}{4}(f+f')\cr
c_{Zf^*f} & = &  -\frac{1}{4}(f\cot\theta_W+f'\tan\theta_W)\cr\cr
c_{We^*\nu} & = & \frac{f}{2\sqrt{2}\sin\theta_W}\cr
c_{\gamma\nu^*\nu} & = & -\frac{1}{4}(f-f')\cr
c_{Z\nu^*\nu} & = &  -\frac{1}{4}(f\cot\theta_W-f'\tan\theta_W)\cr\cr
c_{W\nu^*e} & = & \frac{f}{2\sqrt{2}\sin\theta_W}
\end{eqnarray}

The extension to quarks and strong interactions as well as to
other multiplets and a detailed discussion of the spectroscopy 
of the excited particles can be found in the literature~\cite{pan}.

Here, let us we write down explicitly the interaction lagrangian 
describing the  coupling of
the heavy excited neutrino with the light electron, as it  will 
be used in the following section in order to discuss bounds on the 
compositeness effective couplings from low-energy, nuclear, 
double-beta decay: 
\begin{equation}
{\cal L}_{eff} =  \bigl({ g f \over {\sqrt 2} \Lambda_{\hbox{c}}} \bigr)
\Bigl\{ 
\Bigl( {\overline \nu^*}
\sigma^{\mu \nu} {1-\gamma_5 \over 2} \, e \Bigr)
\, \partial_{\nu} W_{\mu}^+\Bigl\}\, .
\end{equation}

Let us now discuss the current bounds on the mass of the 
excited states derived from fermion scattering experiments.
Limits on $m_{e^*}$ can be obtained from indirect effects due to 
t-channel $e^*$ exchange in $e^+e^- \to \gamma\gamma $.
The L3 collaboration  has found~\cite{adriani} at 
$\sqrt{s} = 91 $ GeV (LEP):
\begin{equation}
m_{e^*} > 127 \, \, \hbox{GeV} .
\end{equation}

Limits on the masses of the excited leptons ($e^* \, , \nu^* $) from
single production in electron-positron collisions $e^+ e^- \to e^* e$,
$e^+ e^- \to \nu^* \nu $  at a center of mass energy corresponding
to the $Z^0$ resonance, are roughly given by  
the $Z^0$ mass\cite{aleph2}:
\begin{equation}
m_{e^*,\nu^*} > 91 \, \hbox{GeV}
\end{equation}
Pair-production of $l^* , q^*$ rely on the electroweak 
charge of the excited particles and gives usually 
less constraining~\cite{pdg} bounds.
These constraints are obviously limited by the center-of-mass energy of
the accelerator.

As regards the bounds on the masses of the excited quarks,
the strongest comes from an analysis of the reaction 
$  p \bar{p} \to q^* X $ with $q^* \to q\gamma , q W$ at
a center of mass energy of $1.8 $ TeV.
Assuming $u^*$ and $d^*$ to be degenerate and 
$f,f',f_s=1$ (where $f_s$ is the 
dimensionless coupling constant of the strong  
transition magnetic coupling $q^* q g$~\cite{pan}, corresponding
to $f$ and $f'$ appearing in Eq.~12)  the
CDF collaboration~\cite{cdf} has found:
\begin{equation}
m_{q^*}> 540 \, \hbox{GeV}.
\end{equation}
The above bounds are summarized in Table~2 and we refer the reader
to ref.~[8] for a list of earlier bounds.
\begin{table}[htb]
\begin{minipage}[b]{\textwidth}
\tcaption{
Current lower bounds on the masses of the excited states of 
ordinary fermions as they are deduced from direct search 
high energy experiments.
}
\label{tab2}
\vspace{5pt}
\small
\centering
\begin{tabular}{||c|c|c||}\hline\hline
{} &{} &{}\\
excited particle & $m_* (\hbox{GeV}) > $ 
& process\\
{} &{} &{}\\
\hline
{} &{} &{}\\
$e^*$  &  127. & $e^+ e^- \to \gamma \gamma $ \\
{} &{} &{}\\
\hline
{} &{} &{}\\
$e^*$ & 91. & $e^+ e^- \to e^+ e^* $ \\
{} &{} &{}\\
\hline
{} &{} &{}\\
$\nu^*$ & 91. & $e^+ e^- \to \nu^+ \nu^* $ \\
{} &{} &{}\\
\hline
{} &{} &{}\\
$q^*$ & 540. & $p \bar{p} \to  q^* +X $ \\
{} &{} &{}\\
\hline\hline
\end{tabular} 
\end{minipage}
\end{table}

Finally, let us remark that the ZEUS and H1 collaborations (DESY)
have recently published~\cite{zeus,hera} 
the results of a search of excited
states from single production in electron-proton collisions
at HERA. They have studied the reaction
$e p \to l^* X $ with the subsequent decay $l^* \to l' V$
where $V= \gamma , Z, W $ (see Fig.~6).
Upper limits for the quantity $\sqrt{|c_{Vl^*e}|^2
+|d_{Vl^*e}|^2}/\Lambda_{\hbox{c}}\times \hbox{Br}^{1/2}(l^* \to l' V)$
are derived\cite{zeus} 
as a function of the excited lepton mass and for the various decay
channels. They are sensitive up to $180$ GeV for $m_{\nu^*}$  and
up to $250$ GeV for $m_{e^*,q^*}$.

For the purpose of comparing  our recent analysis of 
double-beta decay bounds on compositeness\cite{dbd} with
the bounds discussed above, (see section 4)
we quote here the limit on the $\nu^*$ coupling that the ZEUS
collaboration has 
obtained at the highest accessible mass ($m_{\nu^*} = 180$ GeV):
\begin{equation}
\frac{ \sqrt{|c_{W\nu^*e}|^2
+|d_{W\nu^*e}|^2}}{\Lambda_{\hbox{c}}} \times 
\hbox{Br}^{1/2}(\nu^* \to \nu W)\leq 5 \times 10^{-2} \, \hbox{GeV}^{-1}.
\end{equation}

Let us note  that  these limits depend on the branching 
ratios of the decay channel chosen.
\begin{figure}[htb]
\vspace*{4pt}
\begin{center}
\mbox{
\epsfig{file=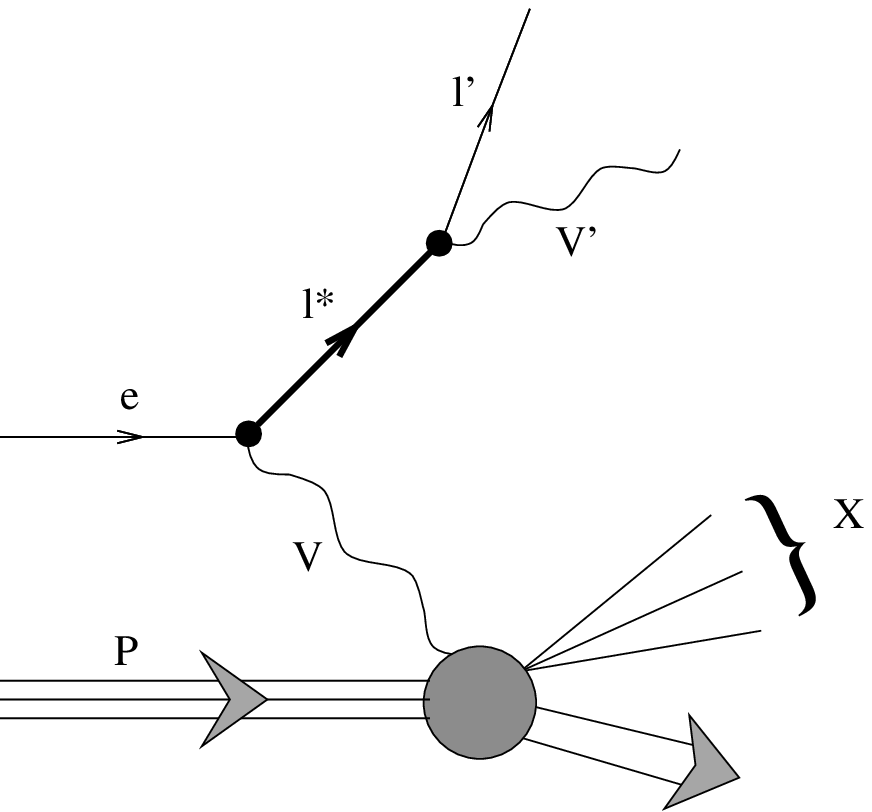,height=7.cm}}
\end{center}
\vspace*{8pt}
\fcaption{
Electroproduction of excited leptons $l^* = e^*, \nu^* $ 
through  t-channel exchange of electroweak gauge bosons
$V = \gamma, Z^0 , W $.
}
\end{figure}

\section{Neutrinoless Double Beta Decay 
(\mbox{\boldmath $ 0\nu\beta\beta $}). }
In this section we discuss the possibility that the heavy 
excited neutrino $\nu^*$ (hereafter denoted $N$) might
be a Majorana particle\cite{dbd,barbieri} and explore
its low energy manifestations, namely neutrinoless double beta decay.

Heavy neutral Majorana particles  with masses in the 
TeV region are predicted 
in various theoretical models,  such as 
superstring-inspired E$_6$  grand unification \cite{e6} or 
left-right symmetric models \cite{lrs}. 
In addition the possibility of a 
fourth generation with a heavy neutral lepton, that could be 
of Majorana
type, is not yet ruled out \cite{hill,datta}. 

In practical calculations of production cross sections and 
decay rates
of excited states, it has been customary\cite{baur,chia,hagi} 
to assume that
the dimensionless couplings $f$ and $f'$ in Eq.~(12) are of order unity. 
However if we assume that the excited
neutrino is of Majorana type, we have to verify that this
choice is compatible with present experimental limits on 
neutrinoless double beta decay  ($ 0\nu \beta \beta$):
\begin{equation}
A(Z) \to A(Z+2) + e^- + e^- 
\end{equation} 
a nuclear decay, that has attracted
much attention both from particle and nuclear physicists
because of its potential to expose lepton number violation.  
More generally, it is expected to give interesting insights 
about certain gauge theory parameters such as leptonic 
charged mixing matrix, neutrino masses etc.
\begin{figure}[htb]
\centering
\vspace*{4pt}
\mbox{\epsfig{file=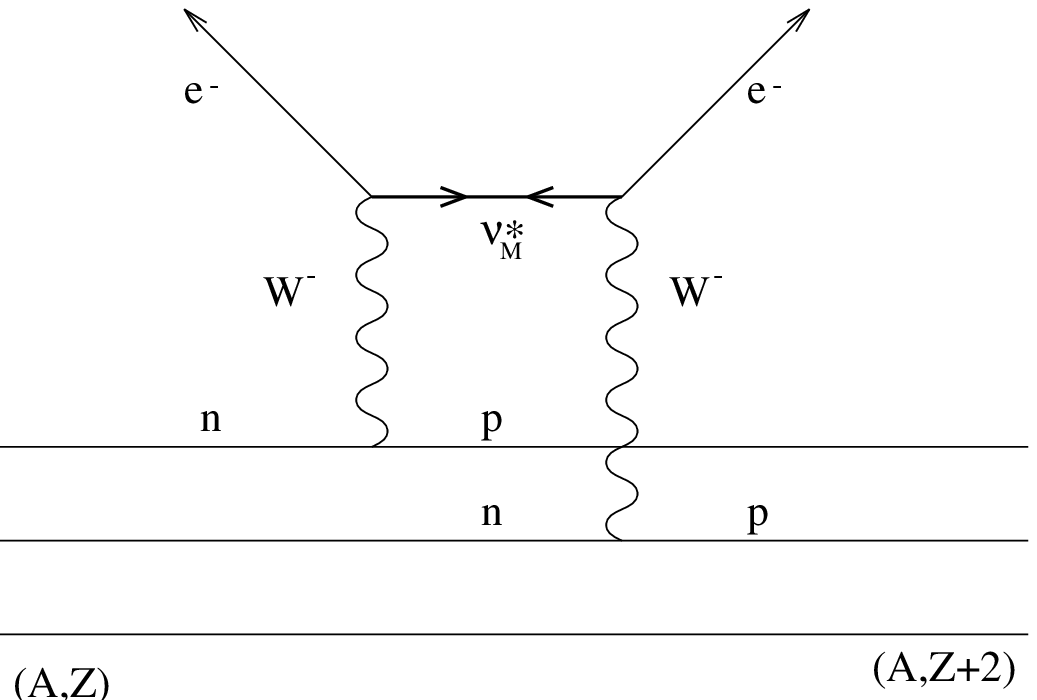}}
\vspace*{8pt}
\fcaption{
Neutrinoless double beta decay ($\Delta L = +2 $ process ) mediated by a 
composite heavy Majorana neutrino.
}
\end{figure}
The process in Eq.~(20), which can only proceed via the exchange of 
a massive Majorana neutrino, has been experimentally searched 
for in a number 
of nuclear systems \cite{exp}
and has also been extensively studied from the 
theoretical side \cite{hax,verga,klapdor}.

We will consider here  the decay
\begin{equation}
^{76}\hbox{Ge} \to \, {^{76}\hbox{Se}} + 2e^-
\end{equation}
for which we have from the Heidelberg-Moscow 
$\beta\beta$-experiment
the recent limit  \cite{balysh} ($T_{1/2}$ 
is the half life =
log2 $ \times $ lifetime)
\begin{equation}
T_{1/2} \,  (\, ^{76}\hbox{Ge} \to \ {^{76}\hbox{Se}} + 2e^-) 
\geq 5.1 \times 10^{24}\, \hbox{yr} \quad 90\,\% \hbox{ C.L.}
\end{equation}

In the following  we estimate the constraint imposed by the above 
measurement 
on the coupling $(f/\Lambda_{\hbox{c}})$ of the heavy composite 
neutrino, as given by Eq.~(15). 
The fact that
neutrinoless double beta decay measurements 
might constrain composite 
models, was also discussed in ref.\cite{barbieri} but 
within the framework of a particular model and referring
to a heavy Majorana neutrino with the usual $\gamma_\mu$ coupling.
Models in which $0\nu \beta \beta$ decay 
proceeds via the exchange of a heavy sterile Majorana neutrino 
(mass in the GeV scale
or higher) have also been recently considered~\cite{burgess}.

The transition amplitude of $0\nu \beta \beta$ decay is calculated 
according to the interaction lagrangian:
\begin{eqnarray}
{\cal L}_{int}& =& {g \over 2 \sqrt{2}}\bigg\{
{f \over \Lambda_{\hbox{c}}}
\bar{\psi}_e(x) \sigma_{\mu\nu}(1+\gamma_5)\psi_N(x)
\partial^\mu W^{\nu (-)}(x)  \nonumber \\
& \qquad & \qquad \qquad  +\cos\theta_C J^h_\mu (x) 
W^{\mu (-)}(x) + h.c. \biggr\}
\end{eqnarray}
where $\theta_C$ is the Cabibbo angle ($\cos\theta_C = 0.974$ ) and
$J_\mu^h$ is the hadronic weak charged current.

We emphasize that in Eq.~(23) 
we have a $\sigma_{\mu\nu} $ type
of coupling as opposed to the $\gamma_\mu $ coupling 
so far encountered in all $0\nu\beta\beta $ decay  calculations
(see the discussion in section 3).
For simplicity, we carry out our analysis
assuming that there are no additional contributions 
to $0\nu\beta\beta$
decay from light Majorana neutrinos, right handed currents
or other heavy Majorana neutrinos originating from another source.

The transition amplitude is then          
\begin{eqnarray}
S_{fi}& =& (cos\theta_C)^2 ({g \over 2 \sqrt{2}})^4 
\left( {f \over \Lambda_{\hbox{c}}}\right)^2 ({1\over 2})
\int\, {d^4k \over (2\pi)^4}\, d^4x\, d^4y e^{-ik\cdot (x-y)}
\times \nonumber \\
&\, &{1\over \sqrt{2}}(1-P_{12})
{\bar u}(p_1)\sigma_{\mu\lambda}(1+\gamma_5)
{\not\! k +M_N \over k^2 -M_N^2} (1+\gamma_5)
\sigma_{\nu\rho}v(p_2) \times
\nonumber \\
&\ & \bigl[ F(Z+2,\epsilon_1)
F(Z+2,\epsilon_2)\bigr]^{1/2} e^{ip_1\cdot x}
e^{ip_2 \cdot y}  \times\nonumber \\
& \ &  (k-p_1)^\lambda(k+p_2)^\rho 
{ \langle f|J^\mu_h(x) \ J^\nu_h(y)|i \rangle 
\over [(k-p_1)^2 -M_W^2] [(k+p_2)^2-M_W^2]}
\end{eqnarray}
where $(1-P_{12})/\sqrt{2}$ is the antisymmetrization operator due to 
the production of two identical fermions, 
the functions $F(Z,\epsilon$)
are the well known Fermi functions \cite{rospri} 
that describe the distorsion of
the electron's plane wave due to the nuclear Coulomb field 
($\epsilon_i$ are the electron's kinetic 
energies in units of $m_ec^2$),
\begin{eqnarray}
F(Z,\epsilon)&=&\chi(Z,\epsilon) 
\frac{\epsilon +1}{[\epsilon(\epsilon +2)]^{1/2}} \\
\chi(Z,\epsilon)&\approx & \chi^{R.P.}(Z)=
\frac{2\pi \alpha Z}{1-e^{-2\pi\alpha Z}}\quad \hbox{(Rosen-Primakoff
approximation)} \nonumber
\end{eqnarray}

As is standard in such calculations, 
we make the following approximations~\cite{hax,verga}:\\
$i$) the hadronic matrix element is evaluated within the closure 
approximation
\begin{equation}
\langle f|J^\mu_h(x) \ J^\nu_h(y)|i \rangle\,  \approx e^{i(E_f
-\langle E_n \rangle)x_0}
 e^{i(\langle E_n \rangle - E_i)y_0} 
\langle f|J^\mu_h(\mbox{\boldmath $x$}) \ J^\nu_h(\mbox{\boldmath $y$})
|i \rangle  
\end{equation}
where $ \langle E_n \rangle $ is an average excitation 
energy of the intermediate states.
This allows one to perform the integrations over $k_0, x_0, y_0$ 
in Eq.~(24);\\
$ii$) neglect the external momenta $p_1$, $p_2$ in the propagators
and use the long wavelength approximation:
$\exp\,(-i{\mbox{\boldmath $ p$}}_1\mbox{\boldmath $\cdot x$})
\, =\, \exp\,(-i\mbox{\boldmath $p$}_2\mbox{\boldmath $\cdot x$})
 \approx 1 $; \\
$iii$) the average virtual neutrino momentum 
$\langle |\mbox{\boldmath $k$}|\rangle \,\,  \approx \, 1/R_0 = 40$  
MeV is much  larger than the typical 
low-lying excitation energies, so that, 
$k_0 = E_f +E_1 -\langle E_n \rangle $
can be neglected relative to \mbox{\boldmath $k$};\\
$iv$) the effect of W and N propagators can be neglected since 
$M_W \approx 80 $\ GeV is much greater 
than \mbox{\boldmath $k$} in the region where
the integrand is large, and we are 
interested in heavy neutrino masses
$M_N \gg M_W $.

For the hadronic current we make the usual ansatz:
\begin{eqnarray}
J_\mu^h (\mbox{\boldmath $x$}) & = &\sum_k j_\mu (k) 
\delta^3 (\mbox{\boldmath $x$} -\mbox{\boldmath $r$}_k )
\nonumber \\
j_\mu(k) & = & {\overline {\cal N}}_k \gamma_\mu (
f_V -f_A\gamma_5)\tau_+(k){\cal N}_k \, f_A(|\mbox{\boldmath $q$}|^2)
\end{eqnarray}
where $\hbox{\bf r}_k $ is the coordinate of the k-th nucleon, 
$\tau_+(k) = (1/2)
(\tau_1(k)+i\tau_2(k) )$
is the step up operator for the isotopic spin, (${\vec\tau}(k)$ is the
matrix describing the isotopic spin of the $k$-th nucleon),
$ {\cal N} = \left ( {\psi_p \atop \psi_n } \right )$
and we have introduced the nucleon form factor:
\begin{equation}
f_A(|\mbox{\boldmath $q$}|^2 )={1\over (1
+{|\mbox{\boldmath $q$}|^2/ m_A^2})^2},
\end{equation}
with $m_A=0.85 $ GeV, 
in order to take into account the finite size of the nucleon,
which is known to give important effects for the heavy neutrino case. 
We also take  the nonrelativistic
limit of the nuclear current:
\begin{equation}
j_\mu(k) =  f_A(|\mbox{\boldmath $q$}|^2)\times\tilde{j}_\mu (k)\, \, 
\phantom{with}
\, \, \tilde{j}_\mu (k) = \left\{
\begin{tabular}{ll}
$f_V\tau_+(k)\, $ & if $\mu=0$ \\
$-f_A\tau_+(k) (\sigma_k)_i\, $ & if $\mu=i$
\end{tabular}
\right.
\end{equation}
($\vec{\sigma}_k $ is the spin matrix of the $k$-th nucleon).
Then using the same notation as in Ref.~[30] we arrive at
\begin{eqnarray}
S_{fi} &=& (G_F\cos\theta_C)^2{f^2 \over \Lambda_{\hbox{c}}^2}  {1\over 2}
2\pi \delta(E_0 - E_1 - E_2) \times \nonumber \\
& \ & {1\over \sqrt{2}}(1-P_{12}){\bar u}(p_1)\sigma_{\mu i}
\sigma_{\nu j} (1+\gamma_5)v(p_2)  
\bigl[ F(Z+2,\epsilon_1)F(Z+2,\epsilon_2)\bigr]^{1/2} \times 
\nonumber \\
& \ & M_N \sum_{kl} I_{ij}\, \langle f|\tilde{j}^\mu(k)\tilde{j}^\nu(l)
|i \rangle
\end{eqnarray}
where $I_{ij}$ is an integral over the virtual neutrino momentum,\\ 
($\mbox{\boldmath $r$}_{kl}=  
\mbox{\boldmath $r$}_k -
\mbox{\boldmath $r$}_l\, , 
r_{kl}= 
|\mbox{\boldmath $r$}_k -\mbox{\boldmath $r$}_l|\, ,
x_{kl}=m_Ar_{kl}) $ 
\begin{eqnarray}
I_{ij}& = & \frac {1} {M_N^2} \int\, 
\frac {d^3 \mbox{\boldmath $k$}}  {(2\pi)^3} \, 
 \frac {(-k_i k_j)}
{(1+|\mbox{\boldmath $k$}|^2/m_A^2)^4}\, 
\exp\,(i\mbox{\boldmath $k\cdot r$}_{kl})  
\nonumber \\
& = & \frac {1}{4\pi} \frac {m_A^4}{M_N^2}\frac{1}{r_{kl}} \left\{
-\delta_{ij} F_A(x_{kl}) +\frac {(\mbox{\boldmath $r$}_k)_i
(\mbox{\boldmath $r$}_l)_j}
{r_{kl}^2} F_B(x_{kl})
\right\} \\
\hbox{with:}&\,  & \,  \nonumber \\
& \, & F_A(x)= \frac {1}{48} e^{-x} \, (x^2 +x)    \nonumber \\
& \, & F_B(x)= \frac {1}{48} e^{-x} \, x^3  
\end{eqnarray}
Note that we can make the replacement\\ 
$\sigma_{\mu i}\sigma_{\nu j} 
\to (1/2)\{\sigma_{\mu i},\sigma_{\nu j}\}
= \eta_{\mu \nu} \eta_{ij} -\eta_{i\nu} \eta_{i\mu} + 
i \gamma_5 \epsilon_{\mu i\nu j} $
since $I_{ij}$ is a symmetric tensor and, with straightforward 
algebra, we obtain
\begin{eqnarray}
S_{fi} & = & M_{fi} \, 2\pi \, \delta (E_0 -E_1 -E_2)\nonumber \\
M_{fi} & = &(G_F\cos\theta_C)^2\frac {1}{4} 
\frac{-1}{2 \pi} \frac{f_A^2}
{r_0 A^{1/3}} \, l \, \langle m \rangle 
\end{eqnarray}  
where we have defined
\begin{eqnarray}
l & =& {1\over \sqrt{2}}(1-P_{12}){\bar u}(p_1)
(1+\gamma_5)v(p_2)
\bigl[ F(Z+2,\epsilon_1)F(Z+2,\epsilon_2)\bigr]^{1/2}\nonumber \\
\langle m \rangle & = & m_e \eta_N \langle f\,|\,\Omega\,|\,i 
\rangle \nonumber \\
\eta_N & = & \frac{m_p}{M_N} m_A^2 
\left ( \frac{f}{\Lambda_{\hbox{c}}} \right )^2 
\nonumber \\  
\Omega & = & \frac {m_A^2}{m_p m_e} \sum_{k\neq l} 
\tau_+(k)\tau_+(l)
\frac{R_0}{r_{kl}} \left [ \left (\frac{f_V^2}{f_A^2}-
{\vec \sigma}_k
\cdot {\vec \sigma}_l \right ) 
(F_B(x_{kl}) -3F_A(x_{kl}))\right. \nonumber \\
& \, &\hbox{\hspace{1.5cm}}\left.  -{\vec \sigma}_k 
\cdot {\vec \sigma}_l\, F_A(x_{kl}) +
\frac {  {\vec \sigma}_k\cdot\mbox{\boldmath $r$}_{kl} \,
{\vec \sigma}_l\cdot\mbox{\boldmath $r$}_{kl} } {r_{kl}^2} F_B(x_{kl})
\right ]
\end{eqnarray}
and $R_0=r_0A^{1/3}$ is the nuclear radius ($r_0 =1.1 $ fm).

The new result here is the nuclear operator $\Omega $ which
is substantially different from those so far encountered in 
$0\nu\beta\beta $
decays, due to the $ \sigma_{\mu\nu} $ coupling of the heavy 
neutrino
that we are considering.
The decay width is obtained upon integration over the density
of final states of the two-electron system
\begin{equation}
d\Gamma =  \sum_{final\, spins} 
{|M_{fi}|^2}\,  2\pi\delta (E_0 -E_1 -E_2) 
\frac {d^3\mbox{\boldmath $p$}_1} {(2\pi)^3 \, 2E_1}
\frac {d^3\mbox{\boldmath $p$}_2} {(2\pi)^3 \, 2E_2} 
\end{equation}
and the total decay rate $\Gamma$ can be cast in the form
\begin{eqnarray}
\Gamma & = & (G_F \cos\theta_C)^4\frac{(f_A)^4 \,m_e^7 
\,|\eta_N|^2} {(2\pi)^5
r_0^2 A^{2/3}} \, f_{0\nu}(\epsilon_0 , Z) \,
|\Omega_{fi}|^2 \\
f_{0\nu} & = & \xi_{0\nu} f_{0\nu}^{R.P.}\\
f_{0\nu}^{R.P.}& = & |\chi^{R.P.}(Z+2)|^2 \frac {\epsilon_0}{30}
(\epsilon_0^4+10\epsilon_0^3 +40\epsilon_0^2+60\epsilon_0 +30 )
\end{eqnarray}
where, $\Omega_{fi} =\, \langle f|\Omega |i \rangle $, $\epsilon_0 $ 
is the kinetic energy of the two electrons in units
of $m_ec^2$, and $\xi_{0\nu}$ is a numerical factor that corrects
for the Rosen-Primakoff 
approximation  \cite{verga} used in deriving the analytical
expression of $f_{0\nu}^{R.P.}$. 
For the decay considered in Eq.(21), we have~\cite{verga}
 $ \xi_{0\nu} = 1.7 $ and $\epsilon_0 = 4. $  
The half-life is finally written as
\begin{eqnarray}
T_{1/2}& = & \frac {K_{0\nu}\, A^{2/3}}
{f_{0\nu}\, |\eta_N|^2 \, |\Omega_{fi}|^2}\\
K_{0\nu} & = & (\log 2) \frac {(2\pi)^5} {(G_F\cos\theta_Cm_e^2)^4}
\frac{(m_e r_0)^2}{m_e f_A^4} = 1.24 \times \, 10^{16} \, \hbox{yr}
\nonumber
\end{eqnarray}

Combining Eq.~(39) with the experimental limit given in Eq.~(22),
we obtain a constraint on the quantity $|f|/(\Lambda_{\hbox{c}}^2 M_N)^{1/2}$
\begin{equation}
\frac{|f|}{(\Lambda_{\hbox{c}}^2 M_N)^{1/2}} 
< \left ( \frac {1}{m_p m_A^2} \right )^{1/2}
\left [ \frac {K_{0\nu} \, A^{2/3}}{ 5.1 \times 10^{24} \, 
\hbox{yr} \times f_{0\nu}(Z,\epsilon_0)}\right ]^{1/4}
\frac {1} {\, \, |\Omega_{fi}|^{1/2}}
\end{equation}

Given the heavy neutrino mass $M_N $ 
and the compositeness scale $\Lambda_{\hbox{c}} $, we only need to evaluate 
the nuclear matrix element $\Omega_{fi} $ to know the 
upper bound on the value of $|f|$ imposed by neutrinoless
double beta decay. 

The evaluation of the nuclear matrix elements 
was in the past regarded as the principal source of uncertainty
in $0\nu\beta\beta$ decay calculations, but
the recent high-statistics measurement \cite{2nu} 
of the allowed $2\nu\beta\beta$
decay, a second order weak-interaction $\beta $ decay, has 
shown that nuclear physics can provide a very good description of
these pheno\-mena, giving high reliability to the constraints imposed
by $0\nu\beta\beta$ decay on non-standard model parameters. 

Since we simply want 
to  estimate  the order of magnitude of the constraint 
in Eq.~(40) we will evaluate the nuclear matrix element
only approximatly.
First of all the expression of the nuclear operator in Eq.~(34)
is simplified making the following replacement \cite{feshbach}
\begin{equation}
\frac { r_{kl}^i r_{kl}^j} {r_{kl}^2}\,  \to \, 
\langle \frac { r_{kl}^i r_{kl}^j} {r_{kl}^2}\, \rangle \, \to \, 
\frac {1}{3} \delta_{ij}
\end{equation}
The operator $\Omega $ becomes then
\begin{equation} 
\Omega  \approx  \frac {m_A^2}{m_p m_e}(m_A R_0)
\sum_{k\neq l} \tau_+(k)\tau_+(l)
\left (\frac{f_V^2}{f_A^2}-
\frac {2}{3} {\vec \sigma}_k
\cdot {\vec \sigma}_l \right )  F_N(x_{kl})
\end{equation}
where $
F_N = (1/x)(F_B -3F_A)=(1/48)e^{-x}(x^2-3x-3)$ 
with $F_B$ and $F_A$ given in Eq.(31). 

Since we are interested in deriving the lowest possible
upper bound on $\vert f\vert $ given by Eq.~(40), let us 
find the maximum absolute value of the 
nuclear matrix element of the
operator $\Omega $ in Eq.(42):
\begin{equation}
\vert \Omega_{fi} \vert \leq\frac{m_A^2}{m_p m_e}(m_A R_0)
\vert F_N(\bar{x})\vert \left\{\frac{f_V^2}{f_A^2}
\vert M_F \vert +\frac{2}{3}\vert 
M_{GT} \vert \right\}
\end{equation}
where $M_F=\ \langle f\vert \sum_{k\neq l}\tau_+(k)\tau_+(l) 
\vert i \rangle 
$ and 
$M_{GT} =\ \langle f\vert  \sum_{k \neq l}\tau_+(k)\tau_+(l)
\vec \sigma_k \cdot
\vec \sigma_l\vert i \rangle  $ are respectively 
the matrix elements of 
the Fermi and Gamow-Teller
operators whose numerical values for the nuclear
system under consideration are  \cite{hax,verga},
$M_F = 0 $ and $ M_{GT} = -2.56$.
Inspection of the radial function 
$F_N $ (for $x \geq 0 $) shows that its maximum absolute value
is  attained at $x = 0$. 
In Eq.~(43) we have evaluated $F_N$ at $x = 2.28\,  
(r_{kl} = 0.5 $ fm). 
This value of $r_{kl}$ corresponds to the typical 
internuclear distance
at which short range nuclear 
correlations become important  \cite{hax},
so that the region $x \leq 2.28 $ does not give 
contributions to the
matrix element of the nuclear operator. 
We thus find: 
\begin{equation}
\vert \Omega_{fi} \vert \leq  0.6 \times 10^3,
\end{equation}
which together with Eq.~(40) gives: 
\begin{equation}
\frac{|f|}{\Lambda_{\hbox{c}} (M_N)^{1/2}} 
\leq 3.9 \, \, \hbox{TeV}^{-3/2}. 
\end{equation}

However, since we have used an upper bound for the nuclear
matrix element (Eq.~(44)), the above should be taken as 
the most stringent upper bound one could possibly get for 
the quantity $|f|/(\Lambda_{\hbox{c}}^2 M_N)^{1/2}$ given the half-life 
measurement quoted in  Eq.~(22).
An exact evaluation of the nuclear matrix element
will give less stringent constraints than those that can be 
derived from Eq.~(45).

\begin{figure}[htb]
\centering
\mbox{\epsfig{file=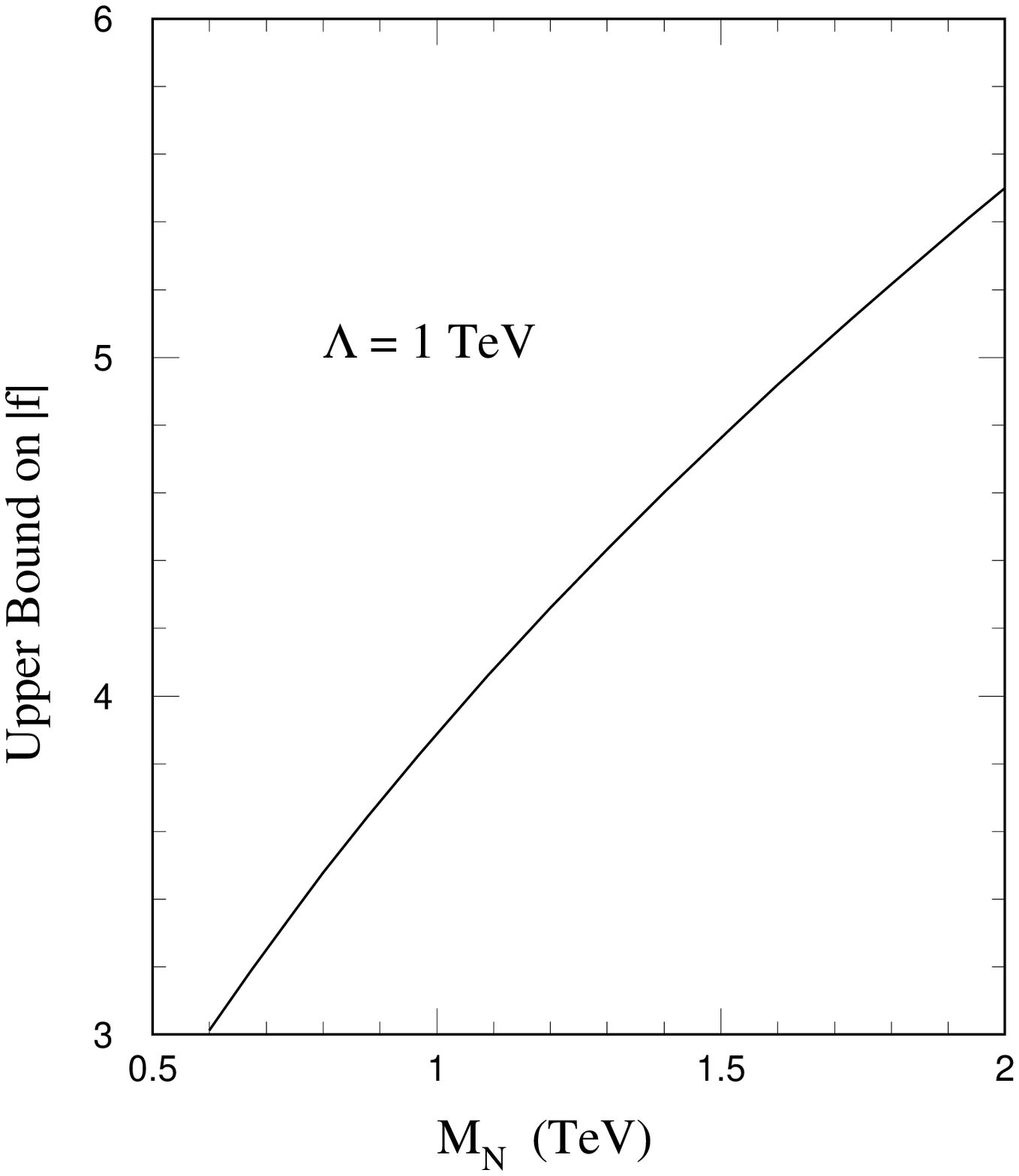,width=10.cm}}
\vspace*{1pt}
\fcaption{
Most stringent ``upper bound'' on $|f|$,
as it can be derived from Eq.~(29),
versus  the heavy Majorana neutrino mass $M_N$ .  
}

\label{boundf}
\end{figure}

\begin{figure}[htb]
\centering
\mbox{\epsfig{file=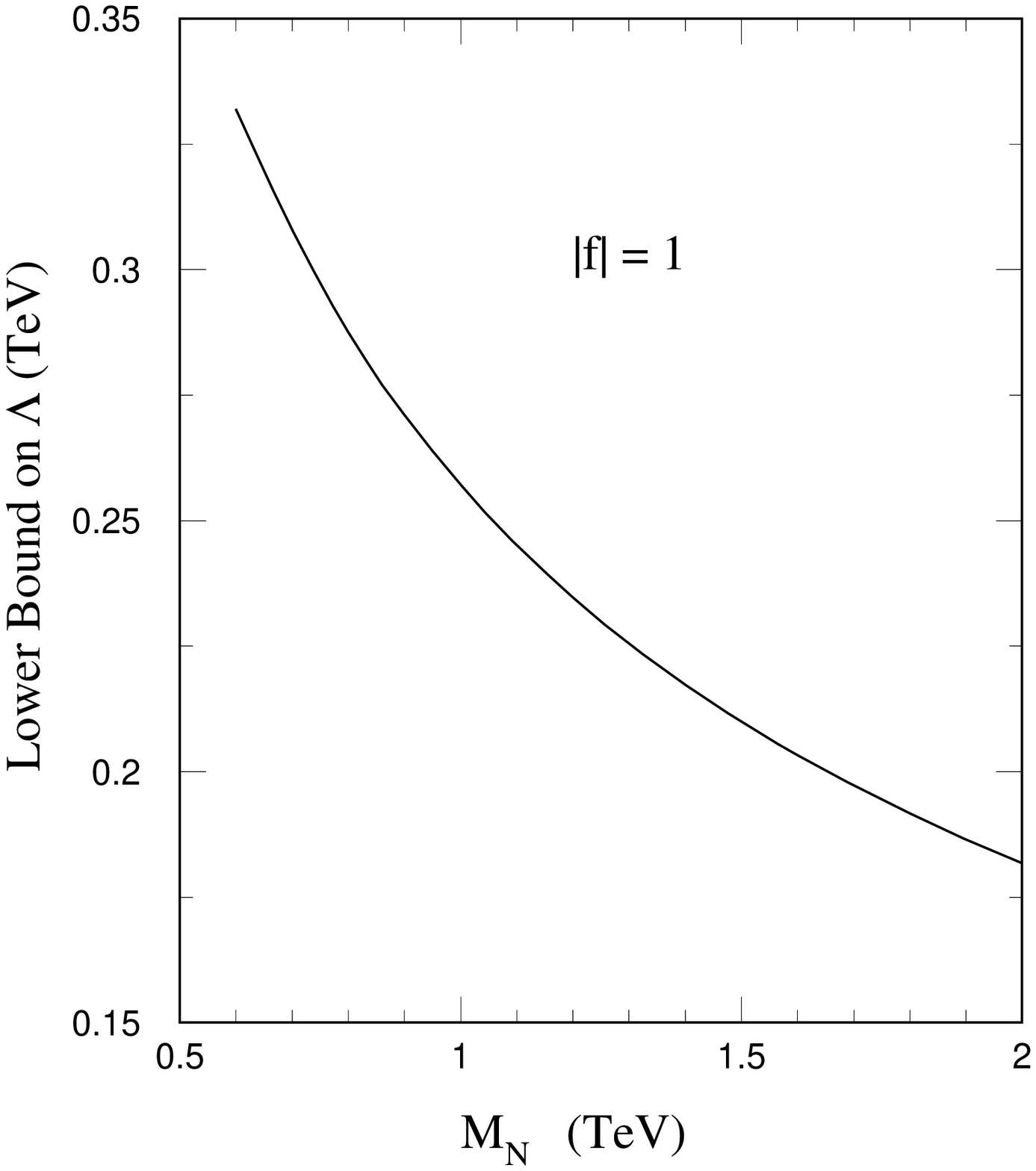,width=10.cm}}
\vspace*{1pt}
\fcaption{
Most stringent ``lower bound'' on $\Lambda_{\hbox{c}}$ versus  
the heavy Majorana neutrino mass $M_N$, as it can be derived 
from Eq.~(29).
}
\label{boundl}
\end{figure}

With this in mind we can use Eq.~(45) 
to give an {\it order of magnitude estimate} of 
the  ``upper bound''
on $|f|$ as a function of $M_N$, 
choosing a value for $\Lambda_{\hbox{c}}$ (See Fig.~\ref{boundf}).
Alternatively, 
Eq.~(45) gives a
lower bound on $\Lambda_{\hbox{c}} $ as a function of $M_N$, 
assuming $\vert f \vert =1 $, (see Fig.~\ref{boundl}). 
We can see that the ``lower bound'' on the compositeness 
scale coming from $0\nu\beta\beta$ decays is rather 
weak: $\Lambda_{\hbox{c}} > 0.3 $ TeV at $M_N = 1$ TeV.

In Table~3 we summarize our bounds for sample values of the excited
Majorana neutrino mass.
\begin{table}[t]
\begin{minipage}[b]{\textwidth}
\tcaption{
Most stringent,
lower bounds on $\Lambda_{\hbox{c}}$ with $\vert f \vert $ = 1, 
and upper bounds on $\vert f \vert $ with 
$\Lambda_{\hbox{c}} = 1$ TeV, for different
values of the heavy neutrino mass $M_N$, as can be derived
from the $0\nu\beta\beta$ half-life lower limit in Eq. (6),
within the approximation discussed in the text.
}\label{tab:bounds}
\vspace{5pt}
\small
\centering
\begin{tabular}{||c|c|c|c|c|c|c|c|c|c||}\hline\hline
{} &{} &{} &{}&{}&{}&{}&{}&{}&{}\\
$M_N \hbox{(TeV)}$ & & 0.6 & 0.8 & 1.0 & 1.2 & 1.4 
& 1.6 & 1.8 & 2.0 \\
{} &{} &{} &{}&{}&{}&{}&{}&{}&{}\\
\hline
{} &{} &{} &{}&{}&{}&{}&{}&{}&{}\\
$\Lambda_{\hbox{c}}\, \hbox{(TeV)} > $ &$[\vert f \vert = 1] $ 
& 0.33 & 0.29 & 0.26 & 0.23 & 0.22 & 0.20 
& 0.19 & 0.18 \\
{} &{} &{} &{}&{}&{}&{}&{}&{}&{}\\
\hline
{} &{} &{} &{}&{}&{}&{}&{}&{}&{}\\
$\vert f \vert < $ &$[\Lambda_{\hbox{c}} = 1 \, \hbox{TeV}] $ 
& 3.0 & 3.5 & 3.9 & 4.2 & 4.6 & 4.9 & 5.2 & 5.5 \\
{} &{} &{} &{}&{}&{}&{}&{}&{}&{}\\
\hline\hline
\end{tabular} 
\end{minipage}
\end{table}
In particular, we see that the choice $\vert f \vert \approx 1 $ is 
compatible with bounds imposed by
experimental limits on neutrinoless
double beta decay rates.
We remark that, as opposed to the case of  bounds 
coming from the direct search
of excited particles, our constraints on 
$\Lambda_{\hbox{c}} $ and $\vert f  \vert $ 
{\it do not depend } on  any assumptions regarding the 
branching ratios
for the decays of the heavy particle. 

Let us now compare our result in Eq.~(45) with that coming
from high energy experiments, c.f. Eq.~(19).
For $m_{\nu^*} = 180 $ GeV (the highest accessible mass at the
HERA experiments~\cite{zeus,hera} with $\Lambda_{\hbox{c}} = 1$ TeV,
$Br(\nu^* \to \nu W)=0.61$~\cite{djouadi} and 
$|c_{W\nu^*e}| = |c_{W\nu^*e}|$ one has:
\begin{equation}
|f|< 61. 
\end{equation}

For the same values of  $m_{\nu^*} = M_N$ and  $\Lambda_{\hbox{c}}$
one obtains from the $0\nu\beta\beta$ constraint i.e. Eq.~(45):
\begin{equation}
|f|< 1.65 
\end{equation}

Due to the approximation in the nuclear matrix element discussed above,
Eq.~(47) represents the most stringent bound that
can be derived from $0\nu\beta\beta$ decay  and  $|f|$ can 
actually be bigger than $1.65 $. 
We can thus conclude that
the bounds that can be derived from the low-energy
neutrinoless double beta decay are roughly of the same order
of magnitude as those coming from the direct search of
excited states in high energy experiments.

To obtain more stringent bounds, we need  to improve
on the measurements of $0\nu\beta\beta$ half-life.
However, our bounds c.f. Eq.~(40) 
on ($\vert f \vert $ or $\Lambda_{\hbox{c}}$ )
depend only weakly on the experimental $T_{1/2}$ lower limit
($\propto T_{1/2}^{\pm 1/4}$). To improve  
the bounds of  an order of magnitude  we need
to push higher, by a factor of $10^4$, the lower bound
on $T_{1/2}$.
We should bear in mind, however, that
the simple observation of a few $0\nu\beta\beta$ decay events,
while unmistakably proving lepton number violation 
and the existence of 
Majorana neutrals,
will not be enough to  uncover the originating 
mechanism (including the 
one discussed here). In order to disentangle the various models,
single electron spectra will be needed, which would require high
statistics experiments and additional theoretical work.

\vspace{0.5cm}
\begin{center}
{\bf Acknowledgements}
\end{center}
This work was partially supported 
by the INFN (Italian Institute for Nuclear Physics), Perugia, Italy.
The author would like to thank the organizers of the 
workshop for their kind invitation, 
C.~Carimalo and Y.~N.~Srivastava
for useful discussions, and                            
the Laboratoire de Physique Corpusculaire,
Coll\`ege de France, Paris,
for its very kind hospitality.

\end{document}